# Disruption Modelling for Engineering and Physics Design of Tokamak Energy's ST-E1 Fusion Power Plant

M. Scarpari[a,*], X. Zhang[a], K. Borowiec[c], P. F. Buxton[a], G. Calabro'[b], S. Carusotti[b], A. Ciula[b], V. Godhani[a], J. D. Lore[c], E. N. J. Maartensson[a], S. A. M. McNamara[a], J. H. Nichols[c], M. Notazio[b], M. Robinson[a], M. Romanelli[a], J. Willis[a], and ST-E1 Team[a]

[a]*Tokamak Energy Ltd., 173 Brook Drive, Milton Park, Oxfordshire, OX14 4SD, United Kingdom.*
[b]*Department of Economics, Engineering, Society and Business Organisation (DEIM), University of Tuscia, Largo dell'Università, Viterbo, 01100, Italy.*
[c]*Oak Ridge National Laboratory, Oak Ridge, TN 37831, United States of America*

**Abstract**

Plasma disruptions represent a critical challenge for high-performance tokamak operations, as they can compromise machine integrity and reduce operational availability. Although future fusion devices essentially need to incorporate strategies to minimise disruption occurrence, complete avoidance remains unattainable. Consequently, assessing and characterising unmitigated disruption consequences is fundamental for the design and qualification of next-generation fusion power plants. This work supports the pre-conceptual design of ST-E1, a low aspect-ratio Tokamak Fusion Power Plant developed by Tokamak Energy Ltd., by presenting a comprehensive disruption modelling approach applied across different design stages. The methodology integrates both physics and engineering considerations to evaluate the impact of disruptions on machine performance and structural integrity. From an engineering perspective, several ST-E1 layout options were analysed to investigate the electromagnetic response of key components under disruption-induced loads, enabling comparison between alternative design solutions. On the physics side, a broad set of disruption scenarios was explored, scanning operational space parameters, plasma-material interactions, and associated thermal loads. Furthermore, the study examined variations in disruption behaviour arising from different reference equilibria, focusing on a range starting from Double Null to Single Null configurations, reflecting the increasing up-down asymmetry consequences. The results reveal significant contrasts in plasma dynamics and structures electromagnetic behaviour between configurations, highlighting the importance of disruption modelling in guiding design choices. These analyses have proven instrumental in shaping ST-E1 development, offering critical insights for mitigating risks and optimising future fusion reactor designs.

## 1. Introduction

Plasma disruptions pose major risks to tokamak operations, threatening both machine integrity and operational availability [1]. Understanding and characterising the resulting ElectroMagnetic (EM) and thermal loads is therefore essential for the conceptual design of future fusion reactors. Disruption generally evolves through two phases [2]. During the Thermal Quench (TQ), the plasma loses confinement and rapidly cools to sub-keV temperatures, producing intense thermal loads on plasma-facing components. The subsequent Current Quench (CQ) involves a fast decay of the plasma current, driven by the sharp rise in plasma resistivity following impurity influx. This decay induces strong eddy currents in surrounding conducting structures, generating substantial EM forces – up to several mega-newtons on the vacuum vessel in JET for current decays over 5-50 ms [3]. The CQ also induces toroidal and poloidal eddy currents due to the Poloidal Field Variation (PFV) and Poloidal Field Toroidal (TFV) [4]. PFV-driven currents, linked to $dI_p/dt$, dominate the axisymmetric EM response in toroidally continuous components such as the vacuum vessel and stabilising plates. In addition, Halo Currents (HCs) represent a second, more localised source of EM loading [5]. When the plasma contacts the wall or divertor during a Vertical Displacement Event (VDE), part of the plasma current flows along open field lines that intersect conducting structures, closing through the vessel and generating intense EM forces. Eddy current induced loads are mainly global and symmetric, whereas halo current loads are localised, transient, and often asymmetric, peaking near the plasma-wall contact regions. Disruptions are typically classified according to the sequence of events leading to the collapse [6][7][8]. Major Disruptions (MDs) occur when the plasma remains vertically centred until the TQ, whereas VDEs arise from loss of vertical stability prior to the thermal collapse. VDEs are generally more critical due to their strong up-down asymmetry, which produces unbalanced EM forces on passive structures.

*author's email: \*Mattia.Scarpari@tokamakenergy.com*

To support the conceptual design of the low aspect-ratio Tokamak Fusion Power Plant (FPP) – ST-E1 [9] – this work presents a 2D axisymmetric electromagnetic modelling approach based on the MAXFEA evolutionary equilibrium code [10]. MAXFEA solves the nonlinear Grad–Shafranov equation using a finite-element formulation and models the plasma evolution with EM coupling to passive and active circuits. The analysis has been applied to two reactor-scale concepts: ST425 (R = 4.25 m) and ST500 (R = 5.0 m) [11], exploring multiple magnetic equilibria – Double Null (DN), Disconnected Double Null (DDN), and Single Null (SN) – to assess the influence of plasma topology on disruption-induced EM loads. The modelling outputs include eddy current maps and EM force distributions over a 2D triangular mesh of toroidally continuous structures, focusing on the Vacuum Vessel (VV) and stabilising plates. The results provide the driver design loads for these components in the ST-E1 pre conceptual design. As additional disruption modelling outcomes, electromagnetic self-consistent plasma motion and thermal loads during vertical displacement phase could be predicted in order to provide a reliable indicator for the design and positioning of limiter protections and, more generally, for the conceptual design of Plasma-Facing Components (PFCs).

A key focus is the comparison between SN and DN configurations. In asymmetric SN equilibria, VDEs produce stronger global vertical forces ($J_{eddy} \times B_{pol}$) than in symmetric DN cases, where up-down balance mitigates the loads [12][13]. Conversely, local halo current effects ($J_{halo} \times B_{tor}$) are more severe in SN VDEs directed toward the X-point region, as observed in EU-DEMO studies [14]. This enhanced severity is linked to (i) weaker vertical field gradients near the null, (ii) absence of up-down symmetry, and (iii) magnetic flux compression in the divertor region, which amplifies local electric fields and current densities [1][15][16]. The present study investigates whether these trends persist at reactor scale in spherical/tight aspect-ratio tokamaks, and whether the SN VDE toward the null represents the worst-case EM loading scenario. It also examines the correlation between magnetic symmetry, flux distribution, and load patterns, supporting the design evolution from ST425 to ST500. Although only unmitigated VDE scenarios were analysed, the results provide a conservative reference for conceptual design [17]. Mitigated disruptions, e.g. through Massive Gas Injection (MGI) or Shattered Pellet Injection (SPI), typically feature slower current decay, smoother beta poloidal ($\beta_p$) evolution, and reduced EM load amplitudes [18][19]. Nevertheless, the overall plasma displacement pattern remains comparable, confirming the relevance of this analysis for defining robust PFCs and limiter protection strategies under worst-case conditions [20].

The following sections of this paper are briefly described as follow: in Section 2 an ST-E1 overview of the project are discussed together with the pre-conceptual design layout description; Section 3 is dedicated to the 2D models description and starting reference equilibria reconstruction for each case analysed; Section 4 deals with the description of the collected disruption cases analysed and the rationale behind those choices, together with the modelling strategies adopted for each case study; Section 5 collect and compare with a final discussion in the conclusion section, results in terms of eddy and halo current distribution, EM forces on ST-E1 layouts key structures and the plasma dynamics during the transient, together with some key features in terms of plasma-wall interaction and correlated thermal loads.

## 2. ST-E1 Overview and Reference Designs

Tokamak Energy Ltd. is developing a pre-concept design of an FPP under the U.S. Department of Energy (DOI) Milestone Program, aiming to demonstrate the engineering and physics feasibility of fusion power based on low aspect ratio and High-Temperature Superconducting (HTS) magnets technology. Two reference cases are evaluated in this work: 4.25 and 5m major radius respectively (see Fig. 1 and Table 1). Disruption modelling activities, performed on both machine scales, have been essential to evaluate how magnetic configuration, machine size, and passive structure layout influence electromagnetic loads and plasma dynamics. The results have provided critical design drivers, particularly for the VV design and the placement and protection strategy of PFCs, supporting the ongoing engineering integration of the ST-E1 concept.

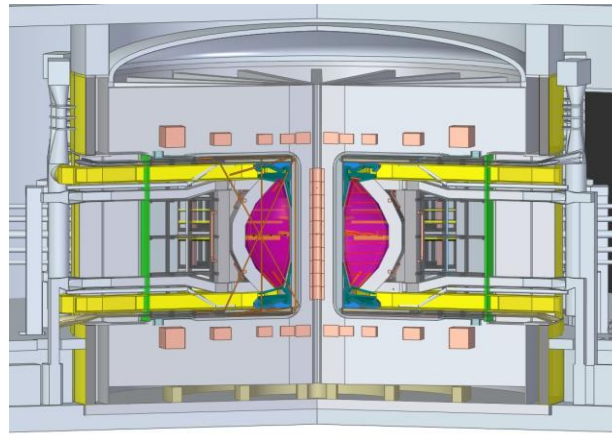

**Fig. 1.** CAD overview of ST-E1 (ST500 reference layout).

The ST-E1 Fusion Power Plant is a tight aspect-ratio tokamak concept developed within the U.S. DOE Milestone Program, integrating HTS magnets. The magnetic system consists of five pairs of Poloidal Field coils (PF) and a ten-segment Central Solenoid (CS), arranged symmetrically around the torus to enable plasma shaping and vertical stability control. Surrounding the plasma, a full-height modular breeding blanket – comprising 36 toroidal sectors – provides neutron shielding and tritium breeding, with helium adopted as the main coolant to avoid strong MHD effects associated with flowing liquid lithium. Vertical stability is ensured through a combination of active feedback coils and passive stabilising structures positioned on the outboard side, which also influence the device's electromagnetic response during disruptions. At the core of the mechanical architecture lies the vacuum vessel, designed as a 50 mm double-shell 316LN grade structure with internal toroidal and poloidal ribs. This configuration offers robust resistance to vacuum, thermal, seismic, and disruption loads while reducing effective toroidal conductivity, thus limiting induced eddy currents during transients. This choice is driven by its proven irradiation tolerance, weldability, and use in fusion-relevant applications. Neutron shielding layers – borated stainless steel or tungsten-boron composites – are integrated within the inter-wall gap to accommodate high mid-plane fluxes, while the vessel is cooled using pressurised nitrogen for compatibility and operational efficiency. Overall, ST-E1 plant architecture is strongly shaped by the vacuum vessel structural, EM, and shielding requirements together with the radial maintenance scheme of the consumable structures, which constrain the arrangement of coils, passive stabilisers, blanket modules and the others in vessel components (such as central column, shielding structures, etc…).

The whole-plant systems modelling presented in [11] identifies a design space which, under the adopted simplified assumptions, yields several design points with comparable overall performance and fully compatible with the high-level mission requirements [9]. All the materials and the correlated properties adopted for the input modelling assumptions are referenced in [9][11][21].

**Table 1.** Major parameters for ST425 and ST500 design points.

|  | ST425 | ST500 |
|---|---|---|
| $R_0$ [m] | 4.25 | 5.00 |
| Aspect-Ratio | 2.15 | 2.30 |
| $B_T$ [T] | 4.0 | 5.25 |
| $I_P$ [MA] | 13.6 | 14.8 |
| $P_{fus}$ [MW] | 870 | 1573 |
| δ / κ | 0.5 / 2.35 | 0.5 / 2.42 |
| $β_N$ | 3.7 | 3.8 |
| Plasma Cross-Section [m$^2$] | 23.9 | 29.2 |

## 3. ST-E1 2D Models Layout Description and Starting Equilibria

This section provides a concise description of the 2D axisymmetric electromagnetic model developed for the ST425 and ST500 configurations. The model includes the main toroidally continuous passive structures, their geometrical layout, material properties, and the initial plasma equilibrium used as the starting point for disruption transient simulations. The modelling was performed using the MAXFEA evolutionary equilibrium code, which solves the nonlinear Grad-Shafranov equation with a finite-element approach, including electromagnetic coupling between the plasma, passive, and active circuit elements.

The simulations start from different flat-top equilibrium reconstruction configuration exploring the differences between SN, DDN and SN reference plasma. The starting equilibrium representative of steady-state plasma conditions immediately prior to a disruption, which is then used as the initial snapshot for the transient disruption analysis. For both ST425 and ST500, the model includes all structures with toroidal electrical continuity (e.g. the VV and stabilising plates), while components lacking toroidal connection, but toroidal geometrical continuity are either represented with a conservative full conductivity or directly neglected when their electromagnetic contribution is marginal.

The PF coils and CS systems are represented according to a non-optimised coil layout but consistent with the early pre-conceptual phase of the ST-E1 design. This assumption likely does not significantly affect the electromagnetic response of the passive structures during disruption transients. Further details on the optimisation of the ST-E1 magnetic cage and PF/CS arrangement are provided in [22].

### 3.1. ST425 MAXFEA Model

The ST425 model corresponds to a ST-E1 design point with a major radius of 4.25 m. The following structures are included (see Fig. 2), with their respective material compositions:

- Vacuum Vessel: Outer shell SS316LN – Intermediate shielding shell SS316L – 80%/WC (central post) – Inner shell SS316LN;
- Divertor Shield;
- Blanket Modules;
- Shield Blanket;
- Central Column;
- High-Field Side (HFS) and Low-Field Side (LFS) Stabilising Plates: SS316LN (no up-down anti-series connection considered at this stage).

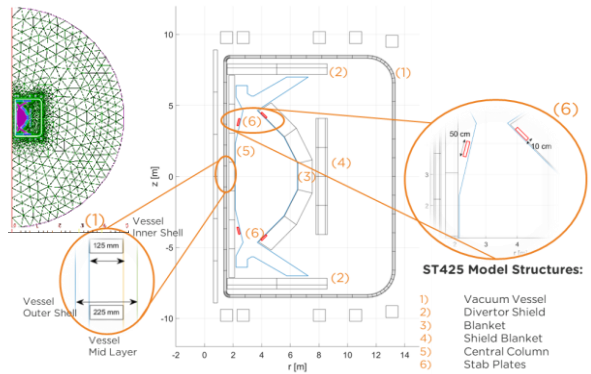

**Fig. 2.** ST425 2D model overview: MAXFEA 2D triangular FEM mesh domain; 2D model profiles with reference structure naming.

The ST425 model includes all structures with toroidal geometrical continuity. For the current analysis, only the vacuum vessel and stabilising plates are considered as passives – all other in-vessel components are replaced by "air" (zero electrical conductivity). This represents the most conservative case for the vacuum vessel that usually provides a reference electromagnetic loads for structural verification at that stage of the design.

### 3.2. ST500 MAXFEA Model

The ST500 model extends the same approach to a larger design point configuration (R = 5.0 m). The structural composition mirrors that of ST425, with updates in geometry and materials for some components with specific focus on VV and stab. plates (see Fig. 3):

- Vacuum Vessel: Outer shell SS316L – Intermediate shielding shell and central post of SS316LN-80% – Inner shell SS316LN;
- HFS and LFS Stabilising Plates: SS316LN (baseline configuration). Alternative cases with tungsten or stainless-steel limiter structures were also investigated as potential passive stability actuators.

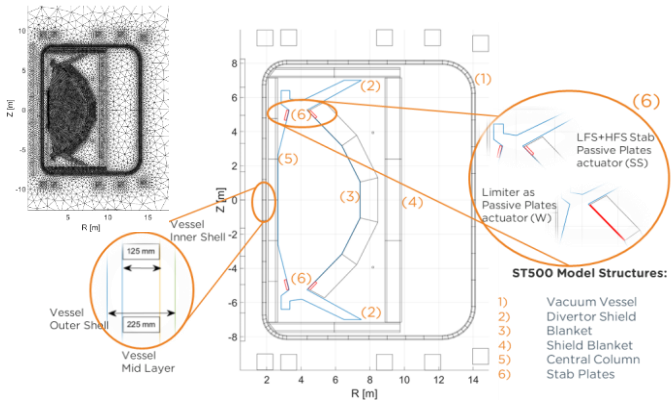

**Fig. 3.** ST500 2D model overview: MAXFEA 2D triangular FEM mesh zoomed to the structures; 2D model geometries with the reference structure naming and with the options for the passive stabilising actuator (SS HFS+LFS stabilising plates or W limiters).

After comparative analysis of induced eddy currents and resulting EM forces, the HFS+LFS SS316LN stabilising plate configuration was selected as the reference option for this study, providing the most conservative load conditions for the vacuum vessel.

### 3.3. Starting Plasma Equilibria Description

The 2D disruption simulations were initiated from a set of steady-state flat-top plasma equilibria representative of the operating conditions immediately preceding a disruption event. The objective was to investigate the influence of different plasma magnetic configurations on the electromagnetic response of power plant structures during VDEs. For the ST-E1 pre-conceptual design, the study of these three configurations allows maintaining design flexibility while assessing their respective advantages and drawbacks in terms of plasma control, exhaust compatibility, and EM load implications.

The equilibria were reconstructed using the MAXFEA free boundary static equilibrium framework, prescribing reference plasma boundaries, current and pressure profiles, and magnet system currents consistent with the ST-E1 reference value [24]. A simplified but optimised set of PF and CS coil currents was applied to reproduce the desired plasma shape and magnetic equilibrium for each configuration, ensuring agreement with the reference design in terms of Last Closed Flux Surface (LCFS), plasma-wall gap, poloidal flux distribution, and magnetic field topology. The reconstructed equilibria are characterised by plasma currents of 13.9 MA for ST425 and 14.8 MA for ST500, with a toroidal magnetic field of 4-5.25 T at $R_0$, poloidal field ∼ 1.2 T at the inboard midplane, $\beta_p$ = 1.4-3.0, internal inductance $l_i$ = 0.5-0.4, and an edge safety factor of $q_{95}$ ∼ 10. In terms of plasma shape the elongation $k$ is around 2.4 and the triangularity $\delta$ is 0.45-0.5. The inter-null separation $\Delta R_{sep}$ varies from 1-1.2 cm in the DDN configuration to > 4 cm in the SN case [24], reflecting the increasing up–down asymmetry. Each equilibrium was validated to ensure physical consistency with the reference design and accurate reproduction of $\psi$-contours, magnetic fields, and LCFS geometry, as required for reliable disruption transient modelling.

These equilibria therefore provide coherent and realistic initial conditions for the 2D electromagnetic simulations, enabling assessment of how configuration topology and magnetic symmetry influence induced currents, EM forces, and the resulting structural response during plasma disruptions. The reconstructed equilibria for the DN, DDN, and SN cases, including the poloidal flux distributions, LCFS locations, and poloidal field magnitude maps, are illustrated in Fig. 4, showing the overall fidelity of the reconstruction with respect to the reference ST500 flat-top plasma condition [24].

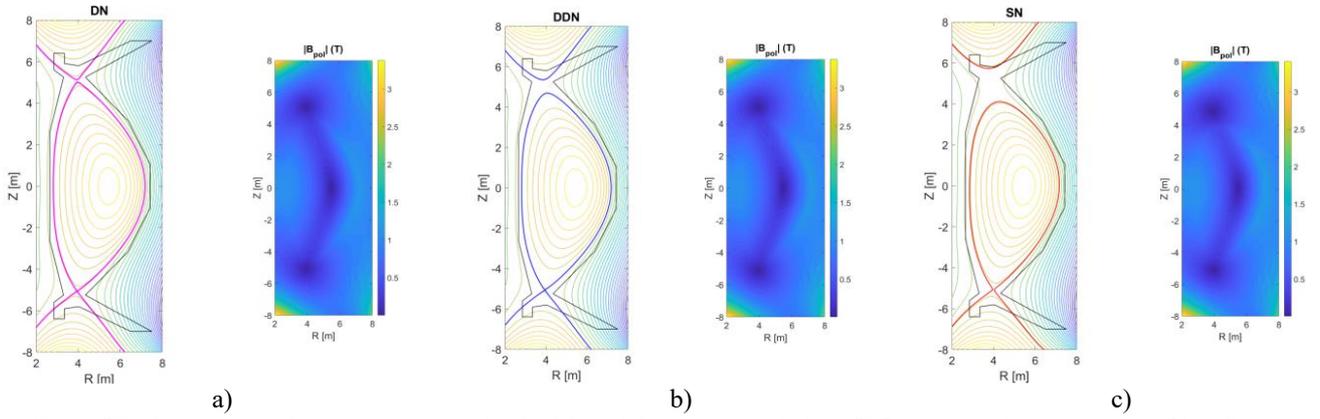

**Fig. 4.** ST-E1 starting equilibria cases. a) $\psi$ and poloidal field distribution with the LCFS represented in magenta for DN reference shape; b) $\psi$ and poloidal field distribution with the LCFS represented in blue for DDN reference shape; c) $\psi$ and poloidal field distribution with the LCFS represented in red for SN reference shape.

## 4. ST-E1 Predictive 2D Disruption Modelling – Methodology

A comprehensive set of disruption scenarios was assembled to establish a preliminary EM load specification matrix for the conceptual design of the main reactor structures – particularly the vacuum vessel – by exploring the effect of different starting plasma configurations (SN, DDN, and DN) on both the plasma dynamic behaviour during the disruption and the induced EM loads on passive components. Each simulation was initiated from the validated flat-top equilibria described in the previous section, using consistent transient modelling parameters and methodology discussed in this section together with a detailed description of each disruption modelling carried out.

### 4.1. ST425/ST500 Disruption Cases

The resulting catalogue of disruption simulations for both ST425 and ST500, summarised in Table 2, thus provides a consistent database of plasma disruptions – structure interaction scenarios and associated EM load metrics, supporting the definition of design-relevant disruption envelopes for the ST-E1 reactor concept. For the ST425 design point, a total of four disruption cases were analysed (see Table 2), primarily aimed at quantifying the differences arising from the three starting equilibrium topologies. In particular, the simulations considered the reciprocal cases of DN upward VDE and DDN/SN downward VDE, reflecting the expected physical behaviour associated with magnetic geometry. In the DN configuration, due to the intrinsic up-down symmetry of both the plasma equilibrium and the machine structure, an upward or downward vertical displacement produces equivalent electromagnetic effects, with nearly identical global and local load patterns. Conversely, for SN and DDN configurations, a VDE directed toward the X-point – typically downward in a Lower Single Null (LSN) or upward in an Upper Single Null (USN) – represents the most realistic and also the most critical case.

This behaviour arises because the magnetic restoring forces are weaker on the null side, and the plasma naturally drifts in that direction during loss of vertical control [16][19][25]. As a result, the X-point-directed VDE generally leads to more severe halo current generation and associated local electromagnetic loads, due to enhanced magnetic flux compression in the divertor region, increased electric fields, and concentrated current paths producing strong $J_{halo} \times B_{tor}$ forces on the vessel and plasma-facing components [16][26][27]. The opposite direction (away from the X-point) typically results in mitigated halo activity and less asymmetric global loads, consistent with multi-machine disruption observations and modelling [14][28].

Among the ST425 cases, both VDEs and Major Disruptions (MDs) were studied. The MD cases, corresponding to approximately symmetric plasma terminations near $Z_c \approx 0$ [8], provide a reference for comparing SN and DN symmetry effects under idealised central disruption conditions, which, although rare, may still occur in large-scale devices [18]. MD results are not reported in this paper. Additionally, two DN configurations were simulated with alternative structural modelling assumptions to bracket the possible electromagnetic load envelope: (i) a vessel-conservative setup, where only the vacuum vessel and stabilising plates were treated as electrically continuous conductors while all other in-vessel components were set to "air" ($\sigma = 0$), representing the most conservative condition for the vessel; and (ii) a fully conducting setup, where all in-vessel passive components were assumed toroidally continuous, representing the most severe load condition for internal structures such as the blanket and divertor shields. Together, these cases define the upper and lower bounds for expected EM forces and currents under disruption conditions in ST425.

For the ST500 design point, a total of five disruption cases were analysed following the same methodology (see Table 2), focusing on SN, DDN, and DN configurations and using the conservative structural option as reference for the vacuum vessel load assessment. In addition to these base cases, three alternative stabilising plate configurations were investigated to assess the impact of different design solutions on eddy current patterns, magnetic energy partition, and resulting EM force transmission to the vessel: (a) the baseline SS316LN plates (disruption modelling reference configuration), (b) SS316LN limiters integrated as stabilising structures, and (c) tungsten limiters acting as passive stabilising elements [24]. These comparisons highlighted the trade-offs between structural load mitigation and current redistribution among passive components, confirming the SS316LN plate configuration as a fair conservative choice for vessel design and a good trade-off in terms of vertical stability options. Proper coupled integrated design of the passive stabilising systems, plasma stability and disruption modelling is left to future work.

Table 2. ST-E1 pre-conceptual design disruption modelling and summary of the cases analysed.

| Starting Equilibrium | Disruption events | DN ST425 Model Structures | DN ST500 Model Structures |
| --- | --- | --- | --- |
| DN | Upward VDE | **Conservative case:** only the vacuum vessel and stabilising plates active (all other in-vessel components as air) → Conservative case for the vessel EM loads. | **Passive stab. option I:** LFS+HFS stab. passive plates actuator in SS316LN |
| Disconnected DN | Downward VDE | **Fully conducting case:** all in-vessel passives electrically continuous → far conservative condition for estimates EM loads on vessel and other in-vessel components. | **Passive stab. option II:** Limiters as passive stab. actuator in SS316LN |
| Lower SN | Downward VDE | - | **Passive stab. option III:** Limiters as passive stab. actuator in W |

**4.2. ST425/ST500 Modelling strategies**

The disruption simulations for ST425 and ST500 were performed using a well-known physically consistent simplified approach reproducing the characteristic phases of tokamak disruption events [4][8][23]. Each VDE simulation reproduces the canonical disruption phases – loss of position control, TQ, and CQ – followed by the growth of halo currents, as already discussed. The VDE modelling methodology and corresponding time-scale assumptions are summarised below:

- **Phase 1 – Onset of vertical motion:** A small perturbation in the radial magnetic field is imposed at the beginning of the simulation to trigger a vertical instability. The sign of the perturbation defines the displacement direction: upward for upper VDEs and downward for lower VDEs. The plasma column subsequently drifts vertically toward the wall under the influence of growing unbalanced EM forces resultant on the plasma.
- **Phase 2 – Thermal Quench (TQ):** The TQ is imposed when the plasma first contacts the wall, marking the transition to limiter conditions. It is modelled as a sharp drop in thermal energy and poloidal beta. The TQ duration ($\tau_{TQ}$ = 0.8-0.9 ms) is assumed similar for both ST425 and ST500 and derived from the ITER Disruption Database (IDDB) scaling [2] with the plasma minor radius (a). Fig. 5b illustrates the empirical scaling of $\tau_{TQ}$ versus minor radius, showing that smaller devices exhibit shorter energy collapse times.
- **Phase 3 – Current Quench (CQ):** After the TQ, the sharp increase in plasma resistivity leads to a rapid decay of plasma current. The CQ duration, defined as the time between 80 % and 20 % of the pre-disruptive plasma current ($\Delta t_{CQ}$ = $t_{80-20}$), is determined by scaling with the plasma cross-sectional area (S) and self-inductance (L*), following the IDDB normalised relation $\Delta t_{CQ/SL*}$ = $\Delta t_{80-20}$ / (S L*) [2]. Fig. 5a presents the distribution of normalised CQ duration $\Delta t_{80-20}$ / (S L*) as a function of toroidal current density ($j_P$) for different devices. This representation enables cross-machine comparison and defines a lower bound on the fastest expected $\Delta t_{CQ}$, represented by the dashed line. Based on this scaling and on the CQ durations adopted are $\Delta t_{80-20}$ ∼ 20 ms for ST425 and $\Delta t_{80-20}$ ∼ 24 ms for ST500, consistent with their respective plasma cross-sectional areas and inductances. During this phase, strong eddy currents and halo currents develop in the surrounding conducting structures, producing both global and local electromagnetic loads.
- **Phase 4 – Halo current evolution:** Once limiter contact occurs, a fraction of the plasma current closes through open magnetic field lines intersecting the conducting structures. The evolution of this halo current fraction and its halo width parameter prescribed are both based on literature observations [2][23].

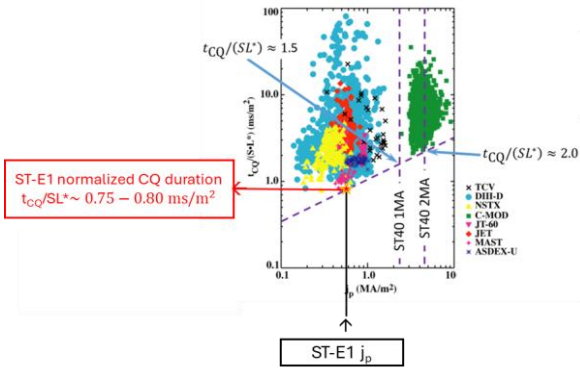

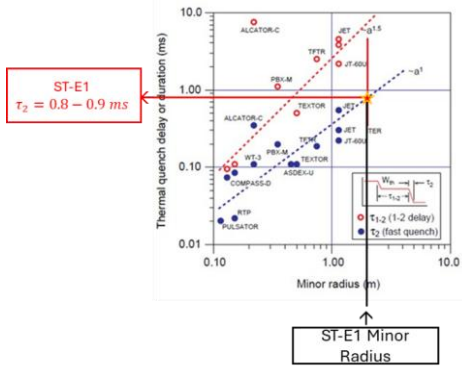

**Fig. 5.** ST-E1 TQ and CQ duration scaling. a) Thermal Quench duration $\tau_{TQ}$ versus plasma minor radius a, derived in [2] and reprinted from [1]; b) Distribution of normalised Current Quench duration versus toroidal current density, separated by device [2], reprinted from [29].

This unified disruption modelling framework ensures physically realistic and conservative transients across all simulated events. Crucially, all disruptions simulated for the three starting equilibrium configurations (DN, DDN, and SN) were modelled using exactly the same event sequence, time scales, and halo current assumptions (e.g. halo region width [23]). This methodological consistency guarantees that any observed difference in electromagnetic response between DN, DDN, and SN configurations arises solely from their intrinsic magnetic topology and plasma geometry, not from variations in model setup – thus providing a robust and directly comparable dataset for the subsequent physical and engineering analysis purpose.

In order to highlight the modelling strategy described above in Fig. 6 – ST425 simulation modelling related and Fig. 7 – ST425 simulation modelling related, are reported on the left the plasma current and halo current numerically obtained from the modelling of the various cases analysed for both the design point, with the timing of the various phases discussed above (TQ/CQ Start and the end of CQ). In addition, a series of snapshot extracted to the main characteristic time during a disruption, such as starting EQ, TQ start instant at the first plasma wall touch, CQ start after the TQ and two snapshots taken at the 80% and 20% of the pre-disruptive plasma current during CQ phase, are proposed on the right in Fig. 6 and Fig. 7.

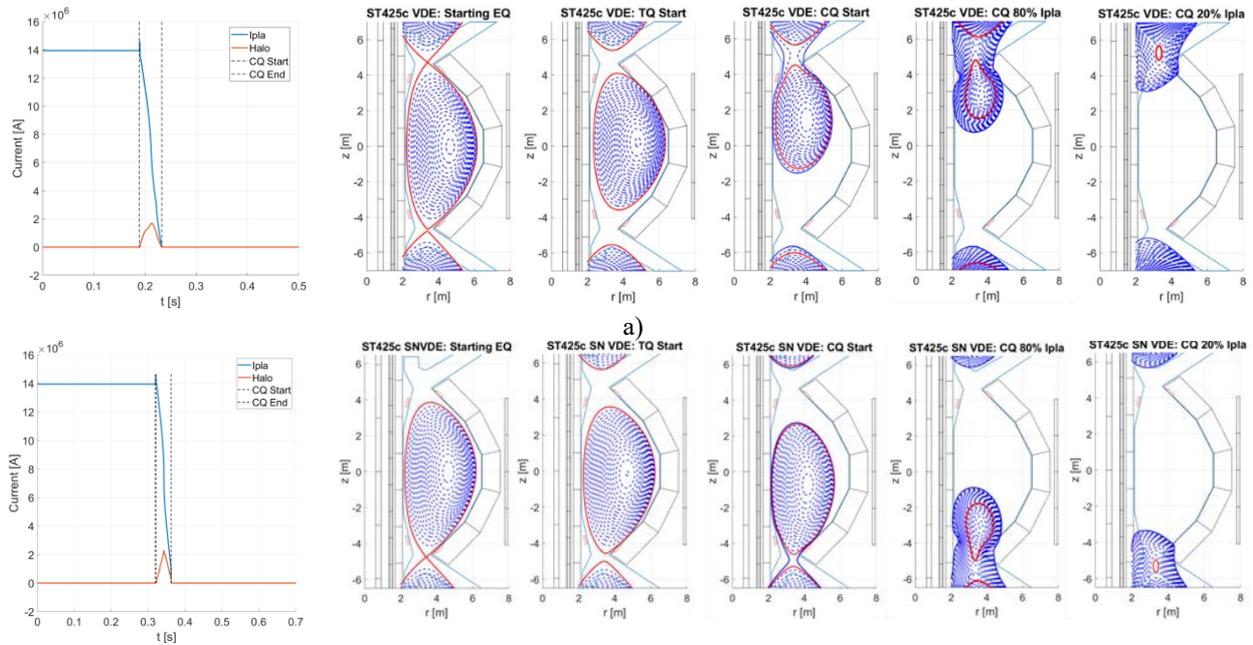

**Fig. 6.** ST425 VDE modelling strategy preview evidence for DN (a) and SN (b): the graph on the left show both plasma current (in blue) and halo current (in red) evolution as direct consequence of the modelling strategies with the timing of the various phases (vertical dashed lines), snapshots sequence on the right report a qualitative dynamic visualization of the plasma shape and halo region evolution.

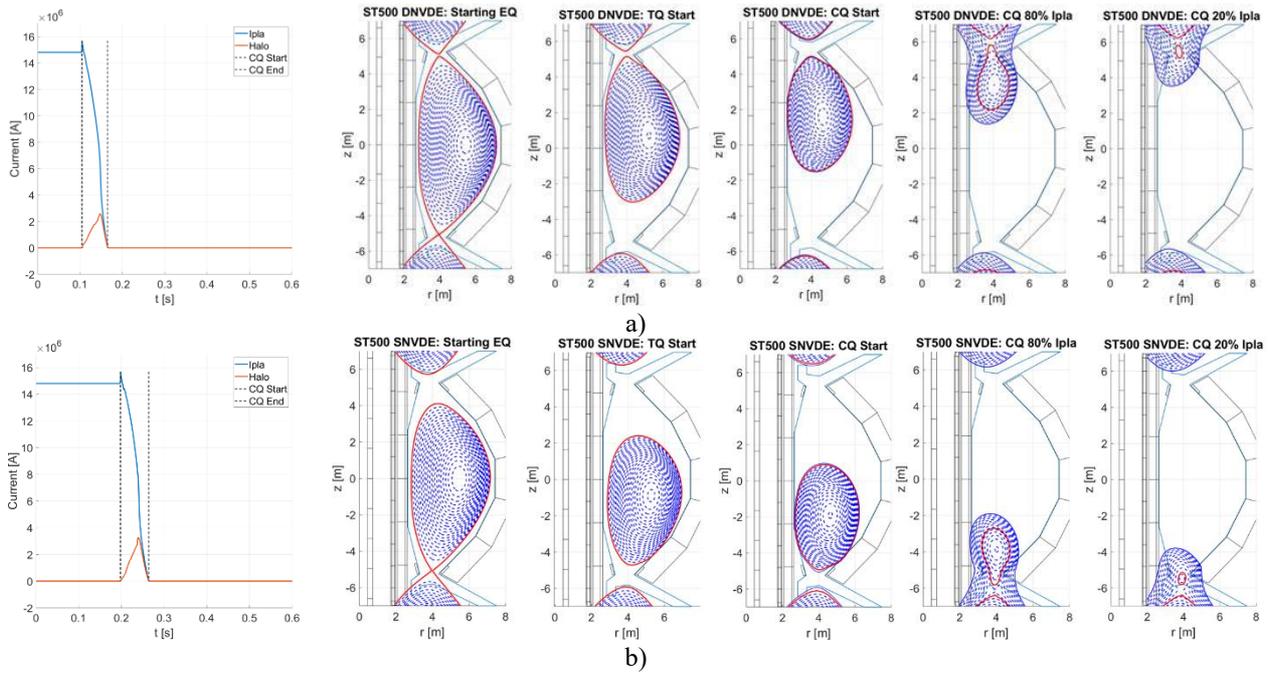

**Fig. 7.** ST500 VDE modelling strategy preview evidence for DN (a) and SN (b): the graph on the left show both plasma current (in blue) and halo current (in red) evolution as direct consequence of the modelling strategies with the timing of the various phases (vertical dashed lines), snapshots sequence on the right report a qualitative dynamic visualization of the plasma shape and halo region evolution.

## 5. ST-E1 Predictive 2D Disruption Modelling – Results

In this section, the main outcomes of the disruption modelling activities performed in the framework of the ST-E1 conceptual design are presented. The objective is to provide a reference set of electromagnetic load drivers for the ST425 and ST500 design options, supporting the assessment of structural performance under disruption conditions. Specifically, the following subsections report the results of all simulated cases described in Section 4.1, including eddy current distributions, global EM forces acting on the main passive components, and plasma dynamic and wall interaction patterns observed during the transients.

### 5.1. ST425/ST500 PFC Disruption Outcomes – Plasma Dynamic Movement and Plasma-wall interaction Evolution

All the main VDE cases with the evaluated plasma movement and dynamic behaviour were collected to determine the plasma trajectory, current redistribution effects, and resulting plasma-wall interaction patterns. Therefore, the evaluation of plasma motion and impact dynamics during disruption events constitutes a key input for the conceptual design of the PFC system, particularly regarding limiter placement and overall layout optimisation. The computed self-consistent trajectories and contact locations were subsequently used as design constraints for defining the limiter geometry and positioning, as it can be possible to observe from Fig. 8 to Fig. 9. All simulated VDE cases were analysed to characterise plasma motion and dynamic behaviour during disruptions, with the aim of identifying the resulting plasma trajectories, current redistribution effects, and plasma-wall interaction patterns. The evolution of the plasma position and its impact points represent a key input for the conceptual design of the PFCs, guiding the limiter placement and the overall layout optimisation of the first wall. The resulting self-consistent trajectories and contact locations, derived from the electromagnetic coupling between plasma and passive structures and subsequent integrations with charged particle heat flux calculation, were subsequently used as design constraints for defining the limiter geometry, as illustrated in Fig. 8 - Fig. 9.

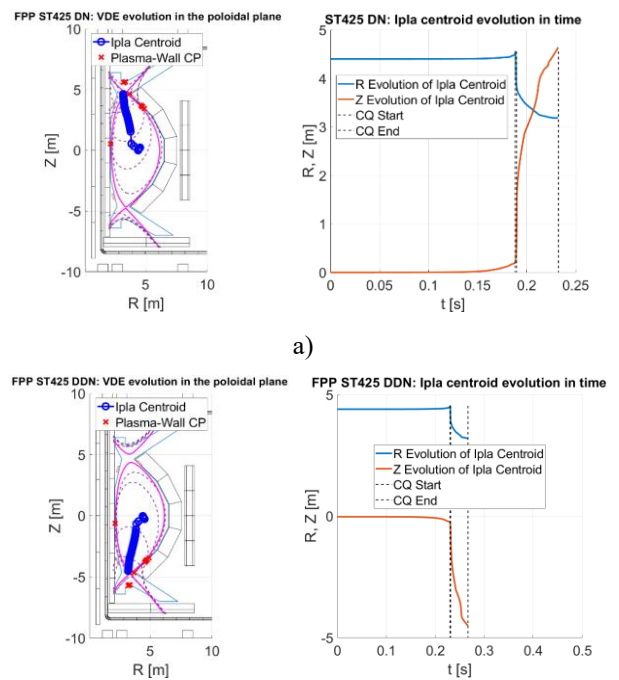

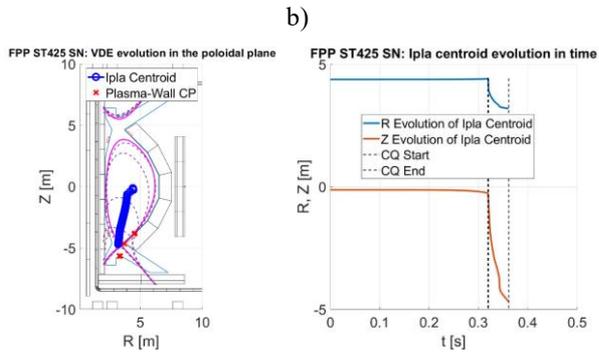

b)

c)

**Fig. 8.** ST425 VDE plasma dynamics for DN (a), DDN (b) and SN (c) presenting the plasma centroid evolutions in time (right figures), on poloidal plane (left figures). In the poloidal plane figures are represented: with the blue trajectory the plasma centroid evolution on poloidal plane, in magenta/dashed purple the plasma boundary evolution during the disruption, "X" red is the plasma-wall contact points in time. In the time variant figures in blue is represented the radial trajectory of the plasma centroid and the vertical one is shown in red.

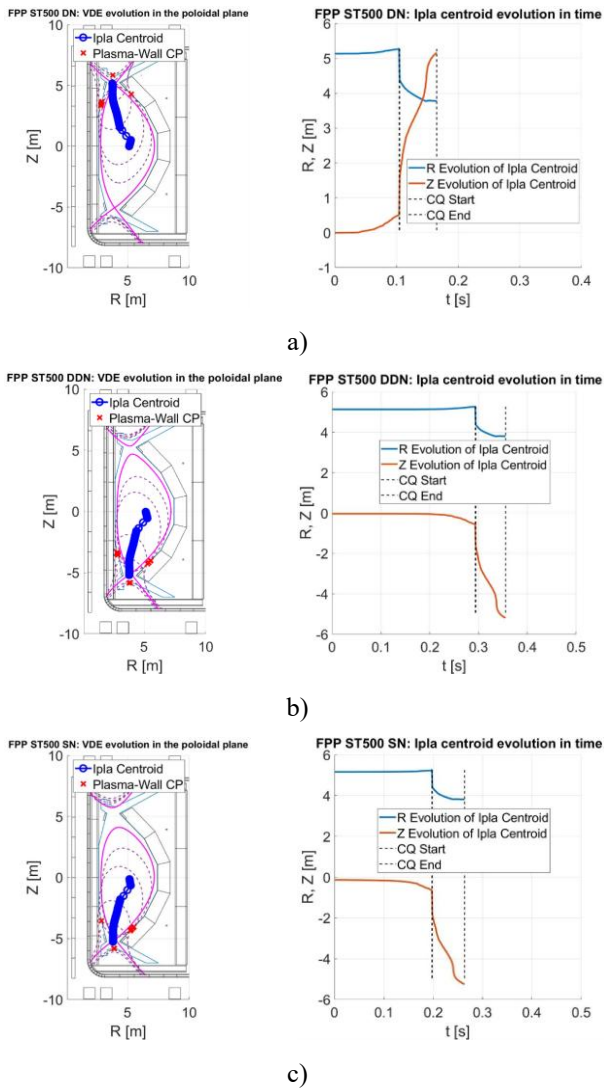

a)

b)

c)

**Fig. 9.** ST500 VDE plasma dynamics for DN (a), DDN (b) and SN (c) presenting the plasma centroid evolutions in time (left figures), on poloidal plane (right figures). In the poloidal plane figures are represented: with the blue trajectory the plasma centroid evolution on poloidal plane, in magenta/dashed purple the plasma boundary evolution during the disruption, "X" red is the plasma-wall contact points in time. In the time variant figures in blue is represented the radial trajectory of the plasma centroid and the vertical one is shown in red.

Fig. 8 and Fig. 9 show, for representative disruption cases and flat-top equilibria, the main dynamic quantities describing the plasma evolution:

- the plasma current centroid trajectory during the VDE (blue circular trace, left-hand side panels);
- the LCFS evolution at key disruption instants – initial equilibrium in solid magenta, first outboard and inboard wall contacts, and two representative CQ snapshots in dashed purple;
- the plasma-wall Contact Points (CP) over time, indicated by red "X" markers; and the radial and vertical motion of the plasma centroid versus time (right-hand side panels, blue and red traces respectively).

Across all analysed cases, the plasma initially drifts toward the south-east/north-east outboard wall during the early displacement phase. Following the imposed TQ, the plasma experiences a rapid inward shrinkage due to the abrupt drop in $\beta_P$, reproducing the macroscopic response to the near-complete thermal energy loss typical of TQs [8][28]. Within a few milliseconds, this collapse leads the plasma to shift contact from the outboard to the inboard wall, given the limited radial spacing between these regions. During the subsequent CQ, the plasma maintains contact with the inboard wall and continues its vertical motion until termination near the divertor dome region – a trajectory clearly observed in the ST500 simulations. In contrast, for the ST425 design, the plasma trajectory exhibits a return toward the outboard sharp corner before proceeding to the dome, consistent with the different vessel geometry and wall curvature. In terms of configuration dependence, no substantial differences were found among SN, DN, and DDN cases for ST500 in the overall trajectory or contact sequence, confirming the dominant role of geometric constraints at this scale. Conversely, for ST425, the SN configuration shows a distinct evolution, with the plasma never contacting the inboard wall, highlighting the influence of magnetic topology and wall shape on the disruption path and plasma-wall interaction pattern.

Fig. 10 (1) provides representative snapshots of the poloidal flux and plasma boundary at two critical instants: the first outboard contact during the TQ and the first inboard contact after the TQ, showing the evolving topology of the plasma-structures interaction and the charged-particle heat fluxes calculation on the sacrificial limiters of ST500 during a VDE.

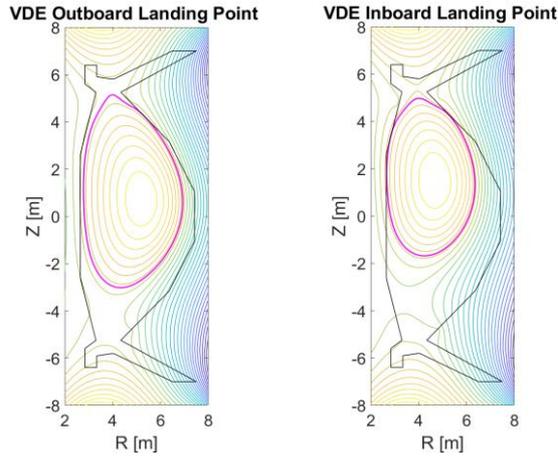

1)

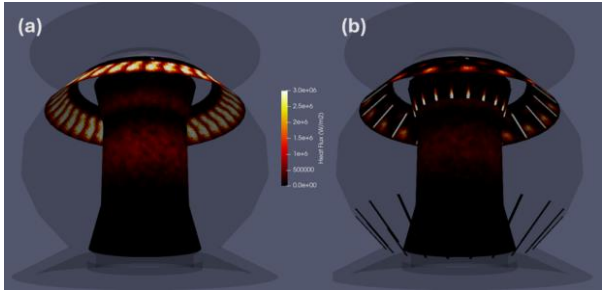

2)

**Fig. 10.** ST500 VDE pre/post TQ plasma-wall collision areas for DN (1). Upward VDE outboard landing point power deposition ($dW_{plasma}/dt$ = 200 MW, $D_m = 3 \times 10^{-4}$ m$^2$/m) in ST500 with a bare wall (2a), or first wall limiter configuration (N=18, d=10 cm) (2b) [30].

Indeed, one of the primary risks to the first wall during a disruption is the deposition of large amounts of plasma stored energy over a very short time interval during the TQ. To evaluate the charged-particle heat fluxes on the sacrificial limiters of ST500 during a VDE, an integrated workflow was implemented between the disruption modelling framework and the DIV3D code [31]. DIV3D is a Monte Carlo field-line diffusion model, recently adapted for tokamak first-wall applications [32]. Although the full spectrum of a disruption scenario is still under investigation, an initial analysis of first-wall heat loads during the DN upward VDE has been performed using DIV3D. Starting from a snapshot immediately before the TQ (Fig. 10 (1), left), the predicted heat flux distribution at the instant when an unmitigated upward VDE first contacts the upper outer wall is shown in Fig. 10 (2). As a first step, an effective steady-state $P_{SOL}$ = $dW_{plasma}/dt$ of 200 MW was assumed, together with a moderate cross-field diffusivity $D_m=3\times10^{-4}$ m$^2$/m. Under these assumptions, the resulting heat flux on a bare first wall is reported in Fig. 10 (2a). The simulation reveals an n=36 modulation associated with the toroidal segmentation of the outer wall [30]. Large areas of the upper outer wall experience heat fluxes exceeding 3 MW/m$^2$; given the limited thermal capability of the breeder wall elements, this scenario would likely lead to widespread damage. Protection limiters can substantially mitigate this failure mode in cases where disruption mitigation is unsuccessful. Applying the same VDE power loading to the limiter configuration discussed in [30] (with N=18 and d=10 cm) yields the heat-flux distribution presented in Fig. 10 (2b). In this configuration, heat fluxes on the breeder wall are almost entirely suppressed, aside from a minor hotspot within each limiter shadow that could be eliminated with more refined shaping. In contrast, each limiter is subjected to heat fluxes exceeding 30 MW/m$^2$. It should be noted that the steady-state input assumption for $P_{SOL}$ may be overly optimistic during a rapid TQ phase, where cross-field thermal transport rises sharply and a large amount of energy is expelled within a very short time frame. For this reason, a more detailed input model and a fully integrated workflow should be adopted in future design steps. Even with this level of detail, a preliminary estimation could be done, and thermal loads are likely to damage one or more limiter modules; however, the modular design allows for rapid replacement. As a result, recovery and return-to-service of the device would be significantly faster compared to scenarios involving widespread first-wall damage.

The modelling outcomes thus provide a reliable indicator for limiter protection design and the definition of robust PFC concepts under worst-case plasma displacement conditions as demonstrated in [20].

### 5.2. ST425/ST500 VV Disruption Outcomes – Eddy and Halo Currents Evolution and EM Loads

This section summarises the main electromagnetic outcomes of the disruption simulations, focusing on the eddy currents, halo currents, and resulting EM forces acting on the reactor structures. All disruption cases introduced in Section 4.1 are included, covering both ST425 and ST500 design points and the three starting plasma configurations (SN, DN, and DDN).

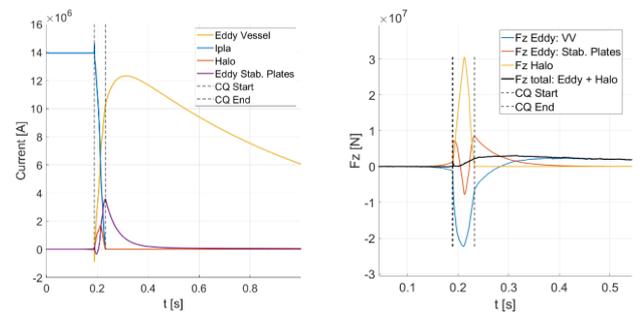

a)

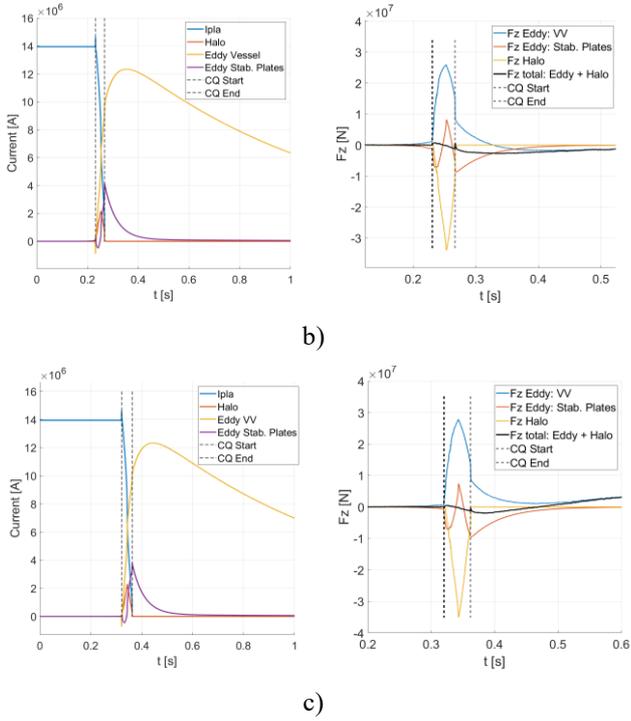

**Fig. 11.** Overview of disruption modelling results for the VDEs ST425 design, including only the VV and stab. plates as passive conducting components. The three rows correspond to DN (a), DDN (b), and SN (c) starting equilibria. Left-hand side panels: temporal evolution of plasma current (blue), eddy currents induced in the VV (yellow) and stab. plates (violet), and halo current (red). Right-hand side panels: corresponding vertical EM forces integrated over the main passive components – VV (blue), stab. plates (red), local halo loads (yellow), and total EM force (black).

Fig. 11 provides an overview of the disruption modelling results for the ST425 design point, considering the reference setup in which only the VV and stab. plates are included as passive conducting components (see Table 2). The three rows correspond to the starting plasma configurations: Fig. 11a DN, Fig. 11b DDN, and Fig. 11c SN. The left-hand side panels show the temporal evolution of the plasma current (blue), together with the eddy currents induced in the VV (yellow) and in the stab. plates (violet), as well as the halo current behaviour (red). When the TQ is imposed, all plasma current trend configurations exhibit a clear current spike associated with magnetic flux conservation and the sudden redistribution of plasma current, consistent with experimental evidence and literature findings [4][8]. The halo currents evolve with a Gaussian-like temporal profile, developing only during the CQ phase, and with a Halo Fraction (HF) around 10-40% (percentage of $I_{halo}$ peak compared to pre-disruption plasma current) as typically observed in literature [2][23]. The right-hand part of the Fig. 11 shows the corresponding evolution of the vertical EM forces integrated over the main passive structures. The blue curves indicate the vertical load on the VV, the red curves the net vertical force on all passive stabilisation system, the yellow curves the local vertical loads associated with halo currents, and the black curves represent the total vertical force, obtained by summing eddy current and halo current contributions.

Quantitatively, the DN upward VDE shows an eddy current maximum on the VV of about 12.3 MA, with a halo current peak of 1.8 MA (HF ≈ 13 %) and a global eddy related vertical force of approximately –22.3 MN. The negative sign of the force indicates a downward-directed EM load on the VV, consistent with the upward plasma motion. The DDN downward case presents a similar eddy level (~12.3 MA) but a higher halo current peak (2.1 MA, HF ≈ 15 %), observing a maximum eddy related vertical load of ~25.9 MN, now upward-directed, following the downward plasma displacement.

The SN downward case exhibits the strongest asymmetry and halo contribution, with a halo current peak of 2.31 MA (HF ≈ 17 %) and a global vertical force on the VV reaching ~27.8 MN from the induced currents contribution, also upward-oriented due to the downward-directed VDE. Overall, the results confirm that while the eddy current response remains comparable across the three configurations, the halo current amplitude and the global vertical EM forces on the vessel progressively increase from DN to DDN to SN, reflecting the higher magnetic asymmetry and enhanced halo coupling characteristic of SN equilibria.

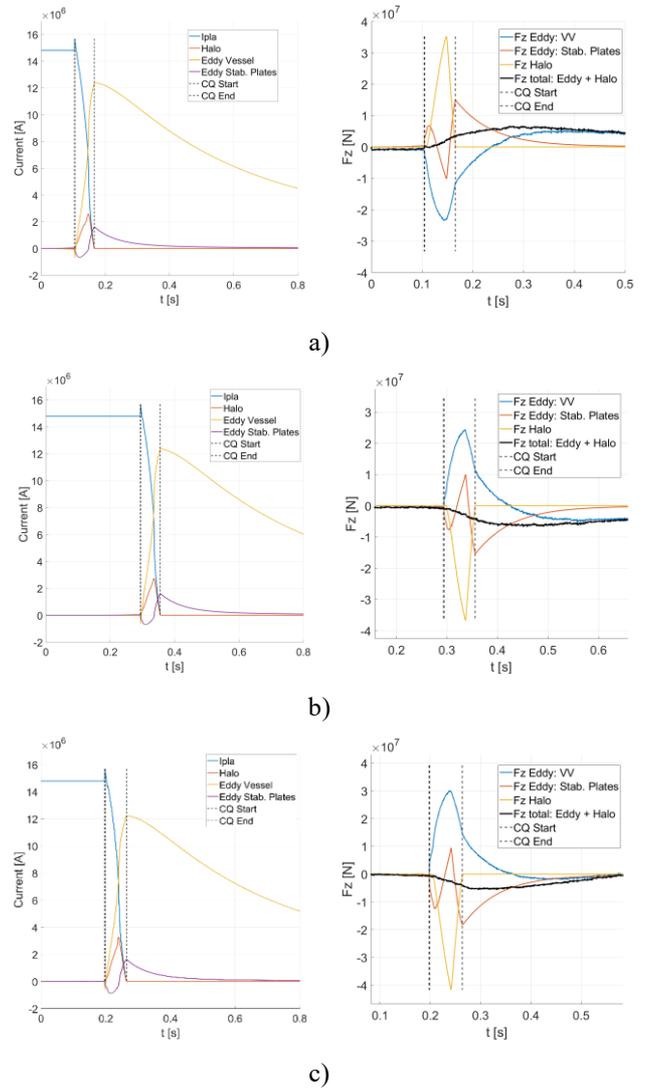

**Fig. 12.** Overview of disruption modelling results for the VDEs ST500 design, including only the VV and stab. plates as passive conducting components. The three rows correspond to DN (a), DDN (b), and SN (c) starting equilibria. Left-hand side panels: temporal evolution of plasma current (blue), eddy currents induced in the VV (yellow) and stab. plates (violet), and halo current (red). Right-hand side panels: corresponding vertical EM forces integrated over the main passive components – VV (blue), stab. plates (red), local halo loads (yellow), and total EM force (black).

Fig. 12 summarises the disruption modelling results for the ST500 design, considering the reference setup with only the VV and SS316LN HFS+LFS (option 1 described in Section 4.1 Table 2) as passive conducting components. The three columns correspond to DN, DDN, and SN starting equilibria. The left panels show the evolution of plasma, eddy, and halo currents, while the right ones report the corresponding vertical EM forces on the main passive structures. The DN upward VDE yields an eddy-current maximum on the VV of about 12.4 MA, a halo current peak of 2.6 MA (HF ≈ 18 %), and a global eddy-related vertical force of –23.6 MN, directed downward as the plasma moves upward. The DDN downward case shows similar eddy levels (~12.4 MA) and a halo peak of 2.8 MA (HF ≈ 19 %), with an upward force of ~24.3 MN. The SN downward case exhibits the strongest asymmetry, reaching 3.29 MA (HF ≈ 22 %) and ~30.1 MN, also upward-directed. The results confirm the trend already observed for ST425 – comparable eddy current magnitudes across configurations, but progressively stronger halo currents and vertical load asymmetries from DN to DDN to SN. Compared to ST425, ST500 shows slightly higher halo peaks and global EM forces, reflecting the influence of machine size and plasma parameters on disruption-induced electromagnetic response, despite a similar modelling approach.

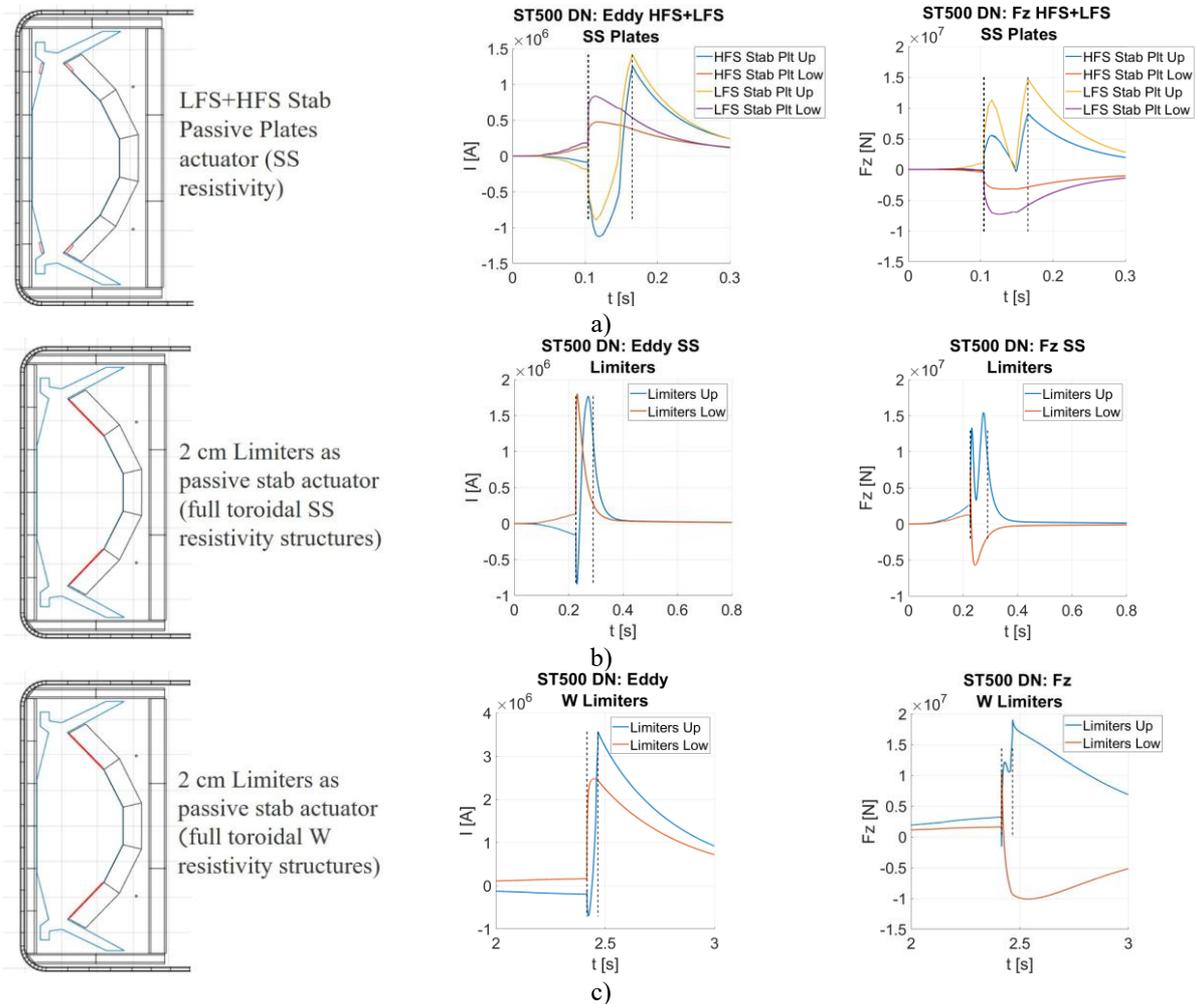

**Fig. 13.** Comparison of three alternative stabilising plate configurations for the DN VDE ST500 setup: (a) baseline SS316LN plates, (b) SS316LN limiters integrated as stabilising structures, and (c) tungsten (W) limiters acting as passive stabilisers. Left: stabilising component layout on the poloidal plane. Centre: induced currents on each passive element. Right: corresponding vertical EM forces (Fz).

Fig. 13 compares three stabilising plate configurations analysed for the DN ST500 case to assess their influence on eddy current distribution, magnetic energy partition, and vertical EM loads on the VV. The configurations are: Fig. 13a baseline SS316LN stabilising plates (reference setup for disruption studies), Fig. 13b SS316LN limiters acting as stabilising elements toroidally, and Fig. 13c continuous tungsten limiters as stabilising elements [33]. Each row of Fig. 13 shows (left) the poloidal layout of the stabilising elements, (centre) the induced currents, and (right) the vertical EM forces on the passive system components. The SS316LN plates generate a maximum vertical load of ~15 MN, while the W limiter configuration reaches ~20 MN, due to stronger EM coupling from their high conductivity and the fully toroidal continuity assumed in the model. The SS316LN limiter option shows intermediate behaviour, offering a balanced compromise between load mitigation and current redistribution. The baseline SS316LN stabilising plates were therefore selected as the reference configuration for disruption analyses, since the detailed 3D design of the limiters is not yet defined at this stage, and real limiters are inherently toroidally discontinuous, thus far from the ideal continuous behaviour of a stabilising structure. Using the SS plates thus ensures a conservative yet realistic assessment of EM loads on the vessel likely maintaining a conservative plasma vertical controllability.

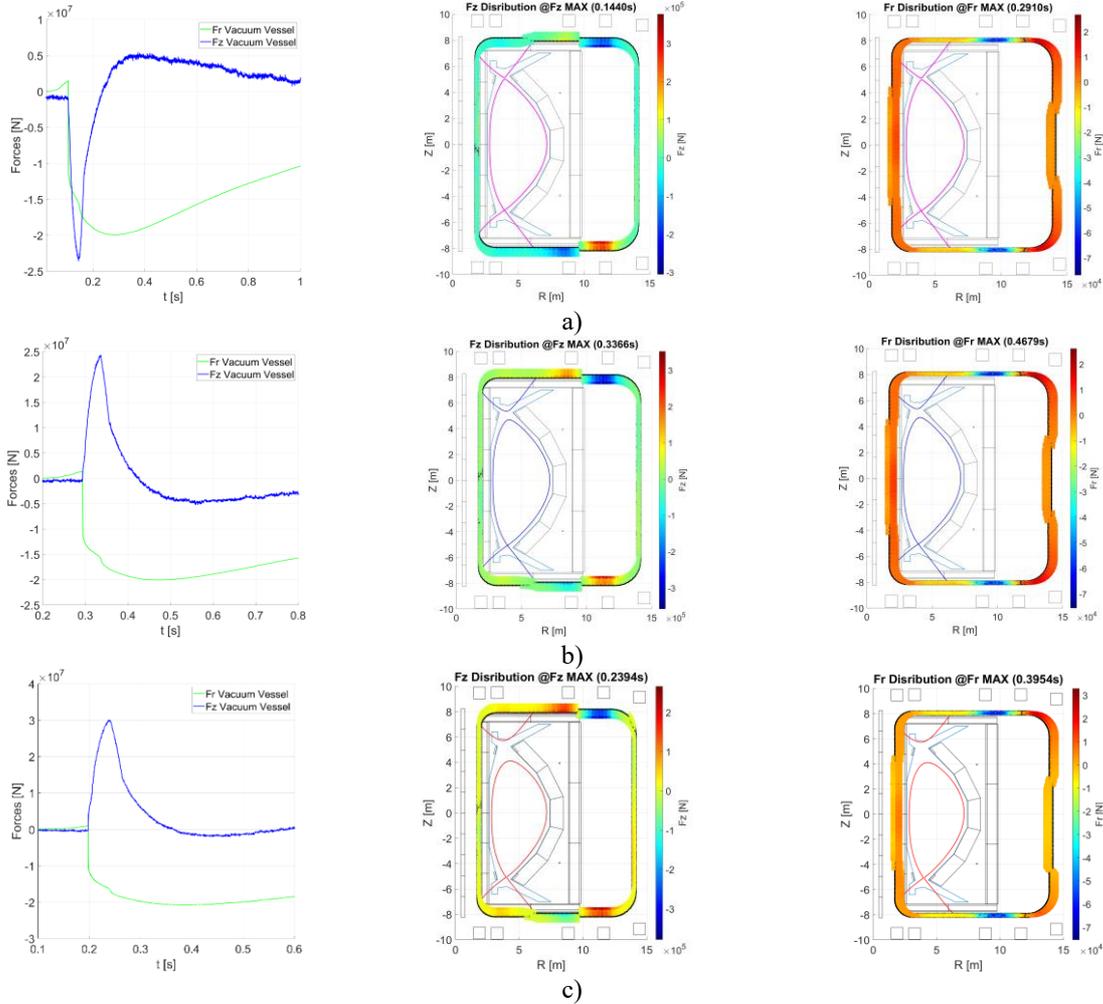

**Fig. 14.** Radial (Fr) and vertical (Fz) electromagnetic loads on the VV evolution of Fz (blue) and Fr (green). Centre: contour–vector maps of Fz at its peak instant – 0.144s for DN, 0.3366 s for DDN and 0.2394 for SN. Right: analogous maps for Fr – 0.2910s for DN, 0.4679 s for DDN and 0.3954 for SN.

Fig. 14 presents the estimated radial (Fr) and vertical (Fz) electromagnetic load distributions on the vacuum vessel for the ST500 disruption cases. The figure is organised in three rows corresponding to the different starting equilibria: Fig. 14a DN, Fig. 14b DDN, and Fig. 14c SN. The left panels show the time evolution of the eddy current induced EM forces, with the Fz in blue and the Fr in green. While the vertical forces have already been reported in Fig. 12, the radial ones are presented here for the first time. Quantitatively, the peak Fr values are approximately –19.9 MN for the DN case, –20.0 MN for DDN, and –20.8 MN for SN, directed inward toward the machine axis. As typically observed in the literature, the radial loads are lower in magnitude than the vertical ones [34] and increase slightly from DN to SN, following the rise in magnetic asymmetry but with a wicker coupling showed by vertical loads. The central panels display contour-vector plots of the vertical force distribution on the poloidal plane at the instant of maximum vertical loads, showing how the distribution varies across the vessel domain. The right panels provide the corresponding radial forces (Fr) maps at the time of their respective peak amplitudes. Both distributions are derived through a post-processing procedure in which the poloidal magnetic field ($B_p$) and current density (j) were extracted from the element centroids of the computational mesh, and the Lorentz force distribution was calculated as $\vec{J}_{eddy} \times \vec{B}_p$ at the centroid of each element of the mesh and integrated on the infinitesimal volume. The resulting $Fr_i$ and $Fz_i$ components were then integrated over the VV domain and in time to obtain the total Fz(t) and Fr(t) trends shown in the left panels. Overall, the results confirm that radial EM loads on the VV remain subdominant with respect to the vertical ones but exhibit a consistent increase with asymmetry, peaking in the SN configuration – in line with the stronger magnetic coupling and eddy current asymmetry observed in previous analyses. Table 3 reports a summary of outcomes in term of global and local EM vertical forces for each design layouts and each magnetic configuration analysed.

**Table 3.** Disruption modelling ST-E1 summary whit focus on global and local effect on the VV for the two design points analysed and for each disruption starting equilibrium.

|  |  | ST425 |  | ST500 |  |
|---|---|---|---|---|---|
| **Starting Equilibrium** | **Disruption events** | **Fz VV Global effect** | **Halo Local effect** | **Fz VV Global effect** | **Halo Local effect** |
| DN | Upward VDE | $Fz_{max}$ ~22.3MN | Halo 1.8 MA | $Fz_{max}$ ~23.6MN | Halo 2.6 MA |
| Disconnected DN | Downward VDE | $Fz_{max}$ ~ 14% more than DN | Halo 14% more than DN in terms of current peak and Fz. | $Fz_{max}$ ~ 3% more than DN | Halo 7% more than DN in terms of current peak and Fz. |
| Lower SN | Downward VDE | $Fz_{max}$ ~ 20% more than DN | Halo 22% more than DN in terms of current peak and Fz. | $Fz_{max}$ ~ 22% more than DN | Halo 21% more than DN in terms of current peak and Fz. |

## 6. Conclusions and Next Steps

As summarized from the disruption modelling performed on the ST425 and ST500 pre-conceptual designs, several consistent trends have been identified regarding the EM loads acting on the main passive structures. Both machines show comparable eddy current induced load magnitudes, with the maximum vertical force on the VV reaching approximately 22.3-27.8 MN for ST425 and 23.6-30.1 MN for ST500, the latter reflecting the stronger coupling and larger scale of the device. These values are of the same order as those reported for other high-field compact tokamaks, although in the ST-E1 designs the VV is located farther from the plasma, and the shielding contribution of the in-vessel components may still be underestimated. The analysis also highlights the importance of vessel geometry and curvature in determining the EM response. The vessel outline and resulting eddy current paths define how efficiently the structure couples with the poloidal magnetic field. In particular, straight horizontal vessel sections near the PFs region enhance the vertical component of $J_{eddy} \times B_{pol}$, producing strong up-down unbalanced forces. Radial loads were also evaluated, showing peak values from 14.9-20.8 MN, all directed inward toward the machine axis. As typically observed in literature, Fr amplitudes are lower than Fz but not negligible, especially in tight aspect-ratio machines, where the strong coupling with the vertical magnetic field amplifies their impact.

Among the analysed plasma configurations, the SN VDE toward the null – identified as the worst-case scenario – produces the most asymmetric and unbalanced vertical forces, driven by the inherent magnetic asymmetry of the SN equilibrium. Correspondingly, halo currents are stronger and more localised than in DN configurations, leading to larger global and localised EM loads. This behaviour is consistent with the central MD simulations, where the SN case again exhibits the highest force imbalance despite geometric symmetry, confirming the dominant role of equilibrium asymmetry. The DDN configuration appears to be a reasonable compromise, offering balanced EM behaviour and design flexibility.

When all in-vessel components are modelled as toroidally continuous conductors, a strong electromagnetic shielding effect is observed: the VV load is largely mitigated, while a significant fraction of the induced currents and forces is absorbed by the Breeding Blanket and Central Column, which directly face the plasma. This configuration provides a highly conservative estimate for in-vessel loads but remains far from the reality, as it overestimates electrical continuity and conductivity, leading to an artificially high redistribution of EM energy within the blanket-shield region. Important EM feedback on the passive stabilising toroidal plates layout design options was also assessed, exploring the possibility to have apposite stab. plates or use the limiter as integrated stabilising actuator system. Proper coupled integrated design of the passive stabilising systems is left to future work. In addition, in the context of PFC design, a preliminary DIV3D analyses show that unmitigated VDEs can deposit multi-MW/m² heat loads over large first-wall areas, posing a significant damage risk for breeder surfaces. Sacrificial limiters effectively localise this power deposition – nearly eliminating first-wall loads at the cost of concentrated heat on modular limiter elements, enabling faster recovery and reduced downtime following a disruption.

Finally, although a 3D electromagnetic analysis – including PFV, TFV, and a comprehensive 3D halo current characterisation – will be required [4] for a refined assessment of local load effects and the assessment of the allowable force, the present 2D axisymmetric modelling provides a robust and conservative basis for the pre-conceptual design phase. In addition, a detailed 3D electromagnetic-to-structural coupling analysis is planned to confirm that the current VV design satisfies the safety criteria and to refine the allowable EM load margins. Indeed, global and local force peaks may exceed material stress limits. For this reason, a full 3D structural analysis is required to accurately determine allowable force limits [35]. The EM load outputs generated in this study have been structured to directly support such analyses in the subsequent design stage. Moreover, even unmitigated disruption scenarios, such as those modelled here, represent a valid and conservative reference for early-stage reactor design, capturing the dominant EM mechanisms and supporting the structural sizing of key components. Future development should focus on extending the analysis to 3D EM modelling and mitigated disruption scenarios (e.g., MGI or SPI) to refine local load prediction and validate the robustness of the conceptual design under realistic operational conditions. In parallel, the disruption modelling campaign has produced complete EM load datasets suitable for direct structural analysis of the VV through a 2D-to-structural fast coupling approach. This method offers an integrated, traceable, and computationally efficient workflow, ideally suited for fast iteration and optimisation during the pre-conceptual design stage of the ST-E1 fusion power plant.

**Acknowledgements**

This work was undertaken as part of the US Department of Energy Milestone-based Fusion Development Program. This work was prepared as an account of work sponsored by an agency of the United States Government. Neither the United States Government nor any agency thereof, nor any of their employees, nor any of their contractors, subcontractors or their employees, makes any warranty, express or implied, or assumes any legal liability or responsibility for the accuracy, completeness, or any third party's use or the results of such use of any information, apparatus, product, or process disclosed, or represents that its use would not infringe privately owned rights. Reference herein to any specific commercial product, process, or service by trade name, trademark, manufacturer, or otherwise, does not necessarily constitute or imply its endorsement, recommendation, or favoring by the United States Government or any agency thereof or its contractors or subcontractors. The views and opinions of authors expressed herein do not necessarily state or reflect those of the United States Government or any agency thereof, its contractors or subcontractors.

**Funding**

Tokamak Energy acknowledges U.S. Federal support for this work under TIA DE-SC0024889. This work is also supported in part by the US DOE under contract DE-AC05-000R22725.